# TrueEBSD in MTEX: automatic image matching for correlative microscopy applications

Vivian Tong[a,1], Stefan Olovsjö[b], Rachid M'Saoubi[b,c], Mathias Grabner[d], Manuel Petersmann[e], Liam Wright[a]

a) National Physical Laboratory, Teddington, TW11 0SN, United Kingdom

b) Seco Tools AB, R&D Materials and Technology, 737 82 Fagersta, Sweden

c) Division of Production and Materials Engineering, Lund University, Olen Römers Väg 1, 221 00, Lund, Sweden

d) Graz University of Technology, Graz, Austria

e) KAI – Kompetenzzentrum Automobil- und Industrieelektronik GmbH, Europastraße 8, 95245 Villach, Austria

1. Current affiliation: Department of Applied Analysis, TU Bergakademie Freiberg, Freiberg, Germany.
Corresponding author email: vivian.tong@extern.tu-freiberg.de

## Abstract

TrueEBSD is an open-source MATLAB program for image alignment and spatial distortion correction of images and electron backscatter diffraction (EBSD) maps. We have re-implemented TrueEBSD as an add-on to MTEX, an established toolbox for EBSD data analysis. Spatial alignment enables correlative analysis methods, such as augmenting EBSD orientation maps with data from other imaging modes. The augmented EBSD maps can then be analysed further using MTEX. We demonstrate TrueEBSD on two example case studies: one for measuring Co phase fraction and WC contiguity in a WC-Co composite, and another for determining the relative susceptibility of grain boundaries to void formation in a copper polycrystal. In both examples, the EBSD map was augmented with scanning electron microscopy (SEM) image data. This enabled quantitative crystallographic measurements which would not be possible from analysing the EBSD maps and images separately.

## Keywords

# 1 Introduction

Correlative microscopy is useful for materials characterisation. Compositing data from different imaging and analysis methods across multiple lengthscales can enable a richer understanding of microstructural features than analysing the data separately. Good spatial registration of the different data types is required correlative microscopy analysis; automating the spatial registration step is particularly important for processing complex experiments and large datasets (such as 3D EBSD), both to reduce operator workload (which would enable larger datasets) and to minimise dependence on operator subjectivity.

'TrueEBSD' is a MATLAB program originally designed to correct spatial distortions in electron backscatter diffraction (EBSD) maps [1]. Distortions between different imaging modes can prevent image or map data acquired from the same physical object from being overlaid pixel-wise, and limit their use in correlative microscopy applications within and beyond scanning electron microscopy (SEM), or large-area analysis using map tile stitching methods [2], [3], [4], [5], [6].

We have re-implemented TrueEBSD using MTEX classes, an open-source MATLAB-based toolbox for EBSD data analysis [7], [8]. This makes it useful for a wider range of use cases and simplifies data pre- and post-processing operations, particularly since MTEX already has an established user base in the EBSD analysis community (https://mtex-toolbox.github.io/). This new implementation of TrueEBSD is released and maintained on GitHub [9] , and supersedes the original MATLAB source code on Zenodo [10].

This update to TrueEBSD has adopted FAIR principles (findable, accessible, interoperable and reusable) in research software development [11]: maintaining the code as a public Github repository (https://github.com/vtvivian/mtex-trueEbsd) improves its **findability**, a modular class structure for handling data increases its **reusability** for a wide variety of experimental use cases, and integration with MTEX 6 improves **interoperability** with various pre- and post-processing methods.

# 2 TrueEBSD-MTEX software

## 2.1 Method overview

TrueEBSD starts with a list of images (which may include orientation or element maps from EBSD or EDS) of the same physical object, but are distorted relative to each other. The program measures and removes these distortions by aligning common features in the images. The distortion-corrected images can be overlaid pixel-wise. The final image of the list is used as the reference image for spatial alignments.

Fast-fourier transform-based image cross correlation of regions of interest (ROI) on the image edge transforms is used to measure a vector field of local shifts between image pairs. These local shifts are fitted to a mathematical model related to the physical distortion type: for example, specimen tilt distortions are be modelled as projective transformations, electric or magnetic field shifts are be modelled as affine transformations, and rolling-shutter distortions from temporal drift can be approximated by a linear spline along the image slow-scan direction. A detailed description of the TrueEBSD algorithm can be found in Reference [1].

Manually selected control points (such as those used in References [2], [12], [13] and short-range corrections in Reference [14]) are not used in TrueEBSD; this allows the process to be fully automated, with the penalty of lower correlation quality near image borders.

## 2.2 Updates to TrueEBSD

TrueEBSD software releases can be downloaded from: (https://github.com/vtvivian/mtex-trueEbsd/releases). This paper describes the code in version 2.1.0.

### 2.2.1 MTEX interoperablility

The new implementation of TrueEBSD is fully interoperable with MTEX 6, and makes use of its EBSD data handling and spatial referencing tools. TrueEBSD inputs and outputs may be pre- and post-processed, visualised and exported using any MTEX and MATLAB functions. Non-image data, such as EBSD orientation maps and elemental maps from energy dispersive X-ray spectroscopy (EDS / EDX), can stored alongside images as a MTEX @EBSD object.

### 2.2.2 Class-based workflow

TrueEBSD has been restructured to use classes[1] for storing image data and metadata such as pixel size and distortion types, and for handling the TrueEBSD workflow steps. This improves the reusability of the program, since having a general template for structuring data enables TrueEBSD to be used for a broader range of applications with minimal code changes.

Three classes have been introduced: the *@distortedImg* class stores a sequence of images and/or EBSD maps and the spatial distortions between them. The *@pairShifts* class stores measured and fitted displacement fields between image pairs, and the ROI size and positions used to calculate these displacement fields. The *@trueEbsd* class stores the data as *@distortedImg* and *@pairShifts* objects at each step of the distortion correction workflow, as well as functions (class methods) to implement each workflow step.

Detailed descriptions of properties and methods for each class, and user guidance on how to optimise tunable parameters, are published and will be kept updated on the mtex-trueEbsd GitHub Wiki page (https://github.com/vtvivian/mtex-trueEbsd/wiki).

The TrueEBSD workflow steps are stored as an object named *job* of class @trueEbsd.

1. *job.imgList* (*@distortedImg* array) contains the initial list of images for alignment. This list starts with the most distorted image and ends with the reference image, and specifies the distortion types between sequential image pairs.
   Any number of intermediate images may be included, and are often useful to separate out compound distortion types: for example, an EBSD map typically includes both tilt and drift distortions. Any image can also include non-image data stored as a MTEX *@EBSD* object.
2. *job.resizedList* (*@distortedImg* array) stores the images after resampling onto a common pixel grid, since the starting images may have different pixel sizes and image extents. *job.resizedList* is a *@distortedImg* array.
   The function *pixelSizeMatch* is used for this step. Images (which can be greyscale, colour or multi-channel) are resampled by linear interpolation, whereas nearest-neighbour interpolation is used to resample non-image data in *@EBSD* objects. Image extents can be aligned either using the top left pixel or central pixel in each image. Images with different extents can be either padded with zeros to match the largest, or cropped to fit the smallest image dimensions.
3. *job.shifts* (cell array of *@pairShifts* objects) stores displacement fields between subsequent image pairs. These include local image shifts calculated between regions of interest (ROI) on the images, pixel-wise shifts computed from fitting the measured image shifts to a specified distortion model, and the residual shifts between image pairs after image correction.

---

[1] A “class” in computer science is a template to describe a set of data and the operations that may be performed on this data. Classes have specified “properties” to store data values, and “methods” to store functions for data manipulation. An “object” is a concrete instance of a class. For example, MTEX has classes defined for different types of crystallography data, and a user may create variables which are objects of these classes. The *@EBSD* class is a template for EBSD map data; a user may create an object named *ebsd* to store and analyse data related to a specific EBSD map. The class properties *ebsd.orientations* and *ebsd.pos* in *@EBSD* are used to access and/or store crystal orientations and map positions.

4. *job.fitError* (*@pairShifts* array) stores displacement fields between subsequent image pairs. These include local image shifts calculated between regions of interest (ROI) on
5. *job.undistortedList* (*@distortedImg* array) stores the images after distortion correction using the pixel-wise shifts, where all spatial features have been aligned to the reference image. *job.undistortedList* is a *@distortedImg* array.

### 2.2.3 Graphical user interface

TrueEBSD now includes a graphical user interface (GUI) option for loading data, which allows the user to inspect data and interactively adjust parameters when setting up the TrueEBSD workflow. With the GUI, no additional coding is required from the user to run simple workflows which require minimal data pre-processing, or advanced modification to the cross-correlation function. Figure 1 shows some example screenshots from the GUI used to load TrueEBSD data, and video demonstrations of the GUI are available online [15].

Figure 1 a shows the image preview tool for loading SEM images from a .h5oina file, an Oxford Instruments HDF5 container file format [16]. This enables the user to preview the SEM images stored in the file and select which ones to use for TrueEBSD, without having to open the file separately in proprietary software supplied by Oxford Instruments such as AztecCrystal.
Figure 1 b shows the image sequence reordering and parameter selection tool which appears once all images and EBSD maps have been loaded into TrueEBSD. The user can reorder the sequence from most distorted (first) to the reference image (last), and remove any data that is not required.
Figure 1 c show the image pixel size input tool which appears when loading images as image files (i.e. not an HDF5 container). The pixel size for each image needs to be provided by the user. In most SEM images, pixel size information is contained in a custom TIFF metadata tag, but this tag is difficult to read automatically because its format and units varies between SEM types and manufacturers. Therefore, parts of the TIFF tag potentially related to image pixel size are shown as a pop-up to provide a hint to the user about the image pixel size required for manual input. The TIFF tag shown in Figure 1 c is an example from a ThermoFisher Apreo SEM.

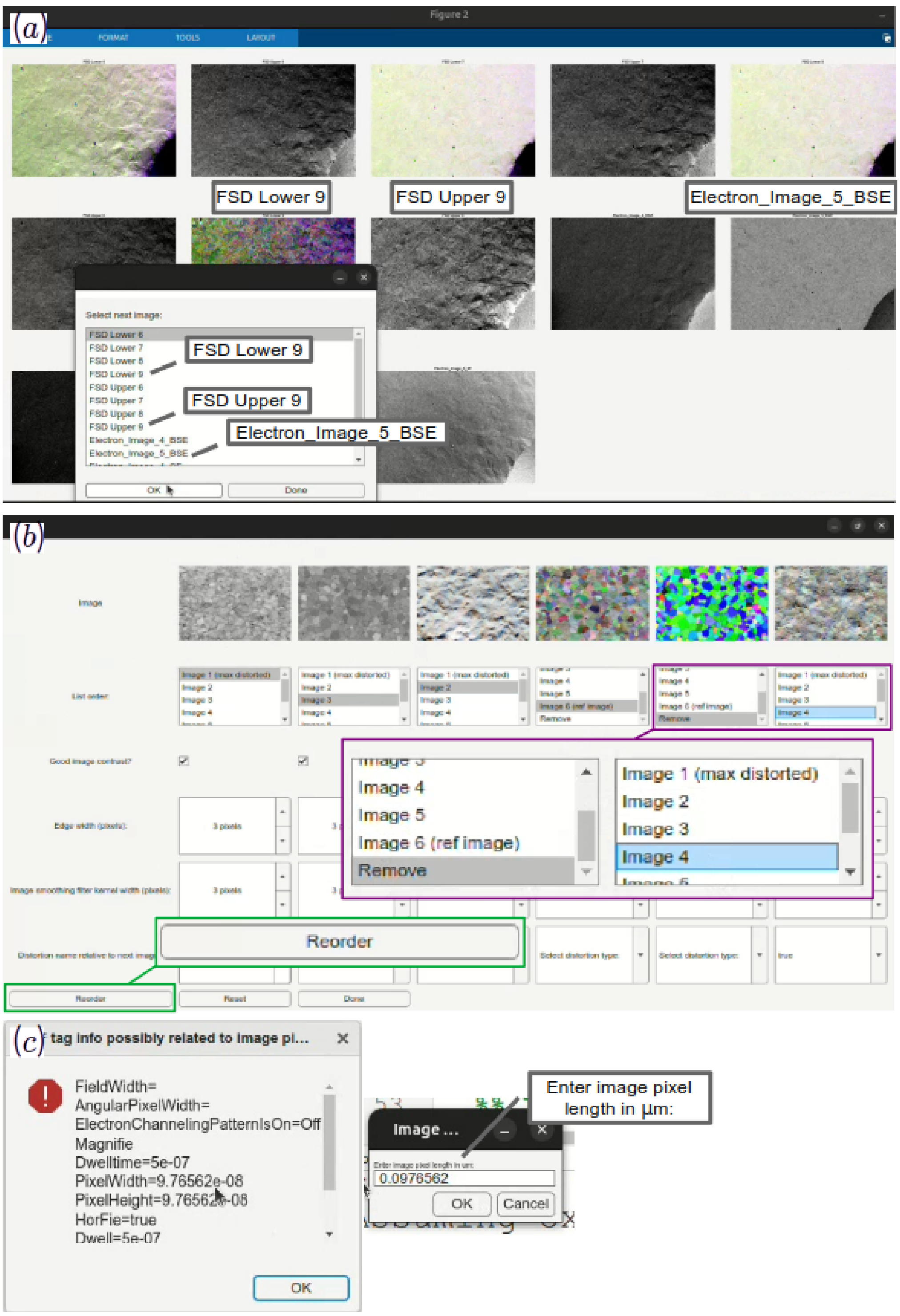

*Figure 1: GUI for data import. (a) .h5oina file image preview tool; (b) TrueEBSD image sequence reordering and parameter selection tool. Outlined boxes show magnified text and portions of the GUI menu; (c) Image file metadata reader and image pixel size input tool.*

### 2.2.4 MEX files

The original TrueEBSD program used MATLAB's Parallel Computing Toolbox (in the *fDIC_xcf_mat* function), since cross-correlation of ROI pairs can be computed independently and therefore suitable for parallel computation.

The new implementation replaces the same function with MEX files (*fDIC_xcf_mat_mex.mex**), which are C binaries generated from the *fDIC_xcf_mat* MATLAB code using the MATLAB Coder Toolbox. This enables significant speedup and still includes parallel computation, and removes the requirement for the Parallel Computing Toolbox.

### 2.2.5 Cross-correlation ROI size

TrueEBSD now iteratively optimises ROI size during cross-correlation, because ROI size is one of the more challenging tunable parameters to optimise.

Larger ROI tend to improve the precision of local image shifts for smooth and slow varying distortion fields. Each ROI should large enough, relative to the image features, to contain sufficient non-colinear features for image registration. ROI should be large enough, relative to the image displacement field, for measured shifts to be no more than around one quarter of the ROI width, because the windowing function applied to the ROI reduces image intensity beyond this [17].

However, increasing ROI size comes at a cost: local shifts near image borders cannot be measured, since every ROI must fit inside the image, and ROI centred near the image border will not fit. The image border 'exclusion zone' is equal to the ROI half-width. This is particularly problematic for EBSD map drift, because drift rate tends to be highest at the start of an EBSD map [12], and its form cannot adequately extrapolated from displacements near the map centre.

The optimal ROI size is the smallest ROI that still allows accurate image registration.

The new TrueEBSD program automatically and iteratively optimises ROI size when calculating image displacements. A small ROI should be chosen to start. If the residual local shifts (*job.fitError*) after image alignment have an average value greater than 2 pixels, this alignment is reattempted with a doubled ROI width, until either the average residual shift is less than 2 pixels, or the ROI width is larger than smallest image dimension.

# 3 Case study: WC-Co phases and WC grain contiguity

## 3.1 Context

WC-Co composites, used in metal cutting applications, can have very fine microstructures that approach the spatial resolution limits of EBSD analysis. Microstructural features such as Co phase fraction, WC grain size, and the spatial distribution of WC grains in the Co binder matrix, affect the mechanical and thermal properties and therefore its performance as cutting tool materials. The degree of contact between WC grains, measured as WC grain contiguity, has been the subject of several studies, but its relationship to the composite structure and material properties is still unclear [18], [19], [20], [21], [22], [23], [24], [25], [26].

A major limitation of EBSD-based characterisation is its asymmetric spatial resolution across WC/Co phase boundaries, because the backscatter yield of the WC phase is greater than Co. EBSD systematically underestimates Co volume fraction, and this error propagates to related properties such as WC grain contiguity. Alternative methods to determine Co fraction include back- and fore-scatter electron (BSE and FSE) imaging, and theoretical binder volume fractions calculated from elemental composition and thermal history using thermodynamic modelling in Thermo-Calc [27].

TrueEBSD was used to overlay and combine WC grain orientations from EBSD maps acquired at 20 kV with WC/Co phase maps from FSE images at 10 kV. The augmented EBSD map was further post-processed to calculate WC grain contiguity from the WC/WC and WC/Co boundary lengths.

## 3.2 Materials and methods

Four WC-Co grades were used as example microstructures, with high and low Co volume fraction, and large and small WC grain sizes respectively. The final specimen surfaces were prepared by argon broad ion beam polishing in a Hitachi IM4000 polisher (5° glancing angle, 2 mm eccentric rotation offset, 3 steps at 6, 3, and 1.5 keV respectively, 30 minutes per step).

Figure 2 shows backscattered electron (BSE) images (10 kV electrons, 0° sample tilt) of the four microstructures.

EBSD maps and FSE images were acquired from a Zeiss Auriga-60 field emission SEM (20 kV accelerating voltage, 120 μm aperture in high current mode, corresponding to approximately 11 nA probe current) using an Oxford Instruments Symmetry 2 EBSD detector. Pixel lengths for all FSE and BSE images were 20 nm for the two 'Coarse WC' microstructures and 10 nm for the two 'Fine WC' microstructures. The EBSD step sizes were 20 nm for the two 'Coarse WC' microstructures, 10 nm for 'Low Co / Fine WC', and 50 nm for 'High Co / Fine WC'.

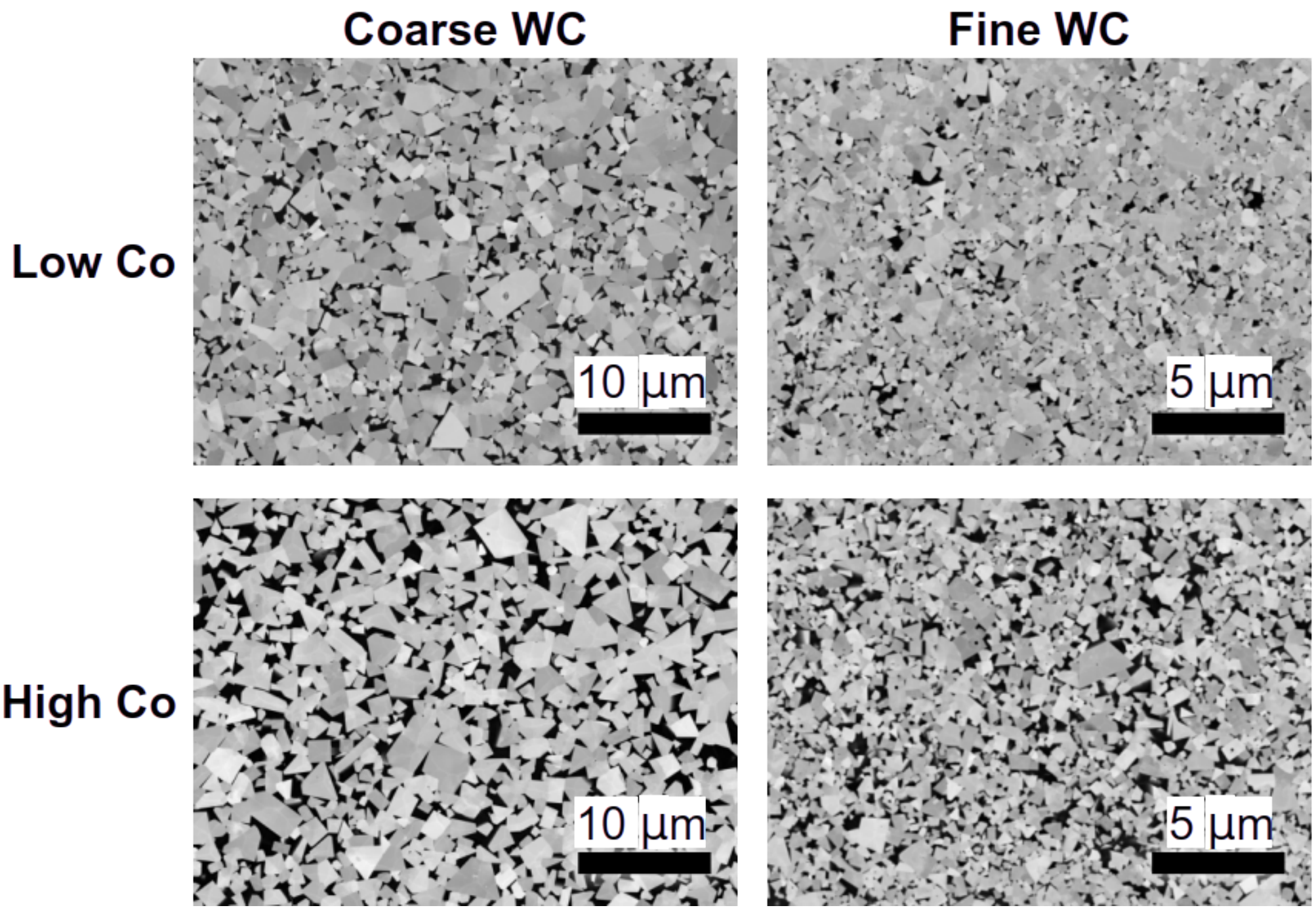

*Figure 2: Example BSE images (0° sample tilt, 10 kV electrons) showing the four WC-Co microstructures used in this work. WC grains (bright phase) are cemented in a Co binder matrix (dark phase).*

Co binder volume fractions were determined using four methods from the same region in each microstructure: EBSD mapping, which can discriminate between phases with different crystal symmetries (hexagonal WC, face-centred cubic (FCC) Co and hexagonal closed-packed (HCP) Co), binary segmentation using Otsu thresholding of greyscale FSE and BSE images, and theoretical predictions using Thermo-Calc. Thermodynamic calculations were performed using the Thermo-Calc software with the TCFE9 database, which, although primarily assessed for steels, contains relevant data for cemented carbide systems. In the calculations, the $M_3C_2$, cementite ($M_3C$) and $W_2C$ (hcp) phases were excluded. Based on Reference [28], atomic mobility is considered to be limited below 1000 °C. Consequently, the binder composition was assumed to become fixed at this temperature. This assumption was applied when calculating phase stability and binder composition at 1000 °C. Table 1 shows how its value depends on measurement method in the four example microstructures. The Co fractions measured using EBSD are systematically lower than both FSE imaging at 10 kV / 70° sample tilt and BSE imaging at 10 kV / 0° sample tilt, which produced comparable results to the ThermoCalc models.

A small fraction of η phase particles were also observed in the Low Co / Fine WC microstructure. The η-phase is a brittle ternary carbide phase, such as $(Co,W)_6C$, that forms under carbon deficiency and is generally detrimental to toughness in WC–Co cemented carbides. In the SEM, η particles show similar contrast to WC in BSE images, but lower (darker) band contrast in the EBSD map. The η particles were mis-indexed as cubic Co in the EBSD map.

*Table 1: Co binder area fractions measured by EBSD, binary segmentation of 10 kV FSE or BSE images, and the theoretical binder volume fraction predicted by Thermo-Calc. The Co binder area fraction for the Low Co / Fine WC material also includes a small fraction of η phase particles mis-indexed as cubic Co.*

| | **EBSD** | **10 kV FSE** | **10 kV BSE** | **ThermoCalc** |
|---|---|---|---|---|
| **High Co Fine WC** | 10 % | 22 % | 21 % | 22.0 % |
| **Low Co Fine WC** | 3.7 % (- η %) | 11 % | 9.3 % | 11.7 % |
| **High Co Coarse WC** | 13 % | 21 % | 21 % | 21 % |
| **Low Co Coarse WC** | 5.9 % | 13 % | 12 % | 10.9 % |

## 3.3 TrueEBSD workflow

The ‘Fine WC / High Co’ microstructure will be used as an example to demonstrate the TrueEBSD workflow in the following section.

Figure 3 shows the starting EBSD maps and images imported into TrueEBSD. The EBSD map (Figure 3 a) was acquired first, and a series of FSE images were acquired after EBSD map acquisition (Figure 3 b - e). Each image was adjusted to approximately match the EBSD map field of view, using the carbon contamination mark after EBSD mapping as a fiducial marker.

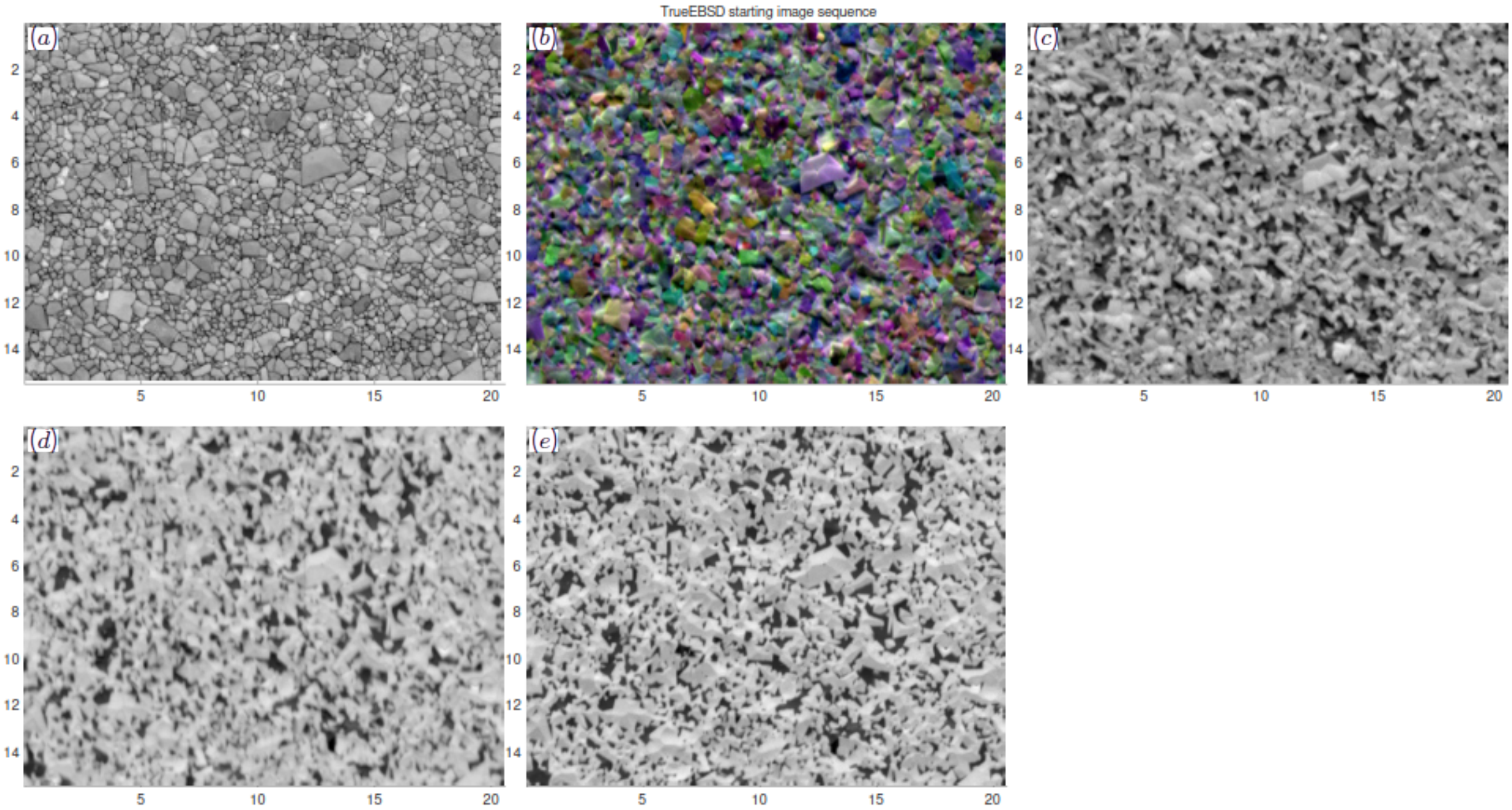

*Figure 3: Imported image sequence. (a) EBSD band contrast map, (b) false-colour FSE image, bottom three diodes, 20 kV, detector retracted 25 mm, (c) FSE image, top two and bottom centre diodes, 20 kV, detector retracted 25 mm, (d) FSE image, top two diodes, 20 kV, detector fully inserted, (e) FSE image, top two and bottom centre diodes, 10 kV, detector fully inserted. Coordinate axes show image scales in μm.*

Figure 4 shows the same figures pre-processed for image alignment. The EBSD map, which had a 5 × longer step size than the pixel size of the FSE images, was upsampled using nearest-neighbour interpolation. Edge transforms of the EBSD band contrast map and false-colour FSE bottom-diode images (Figure 4 a, b) were used to generate similar features for image alignment, whereas the three FSE top-diode images (Figure 4 c, d, e) were aligned using the images themselves. The distortion types between sequential image pairs are listed in Table 2. The FSE top- and bottom-diode images (Figure 4 b, c) were acquired in the same scan. The change in SEM accelerating voltage from 20 kV to 10 kV (Figure 3 d, e) was treated as a tilt distortion because the distortion field incidentally fits very well to a projective tilt transformation, but is not a true projective distortion because the sample did not physically move relative to the electron beam.

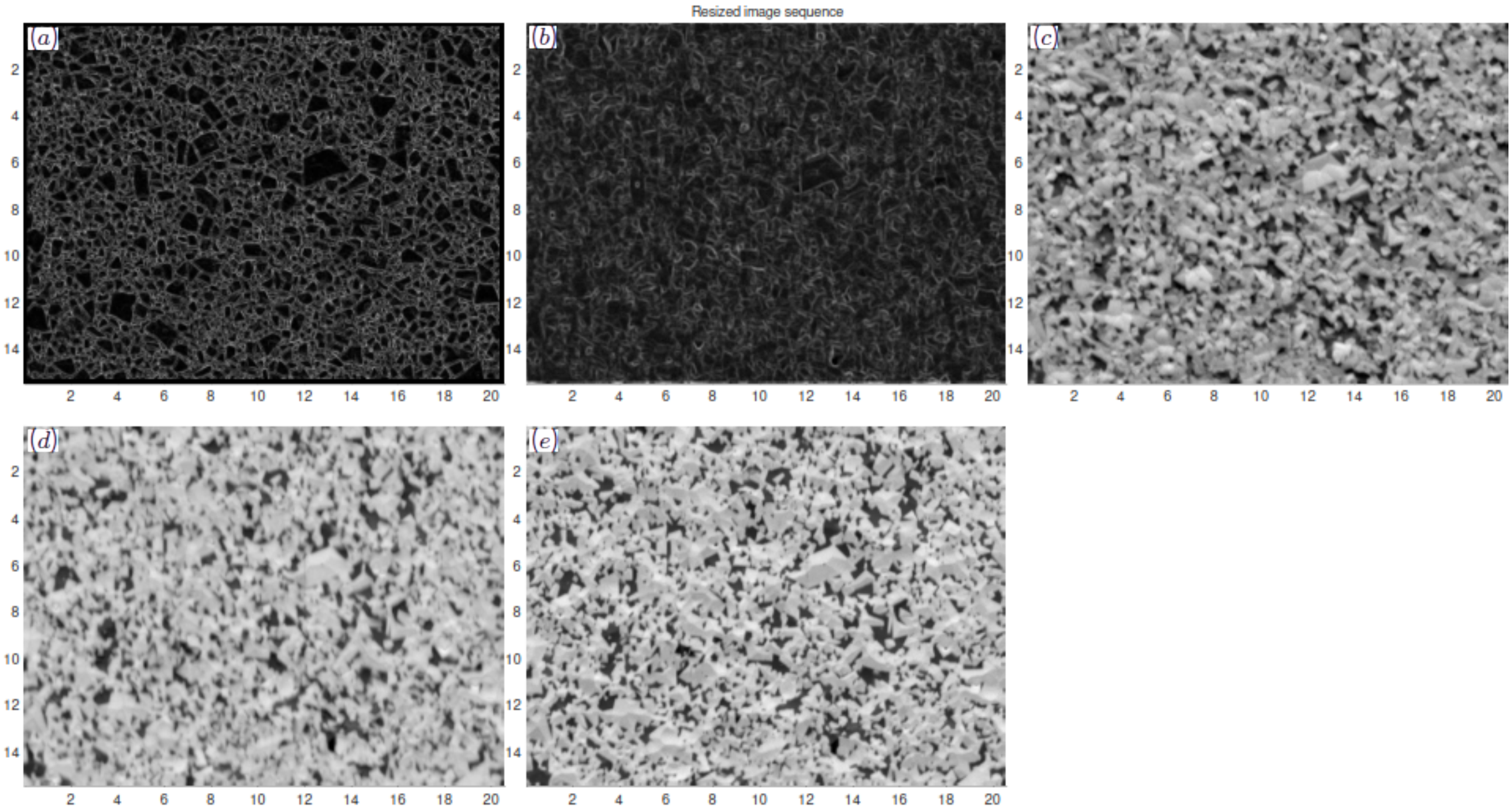

*Figure 4: Pre-processed image sequence. The edge transforms of (a) the EBSD map and (b) the FSD bottom-diodes image have shared contrast for image matching even though the original images (Figure 3 a, b) do not. Coordinate axes show image scales in μm.*

*Table 2: Physical origins of distortion types between sequential image pairs and mathematical distortion models used to fit a continuous displacement field to correct these distortions.*

| Image | Distortion model (type) | Physical origin |
|---|---|---|
| EBSD band contrast map | Affine transform + linear interpolation between rows (drift-shift) | DC electromagnetic field shift from camera movement + temporal drift |
| 20 kV FSE image, bottom diodes, camera retracted 25 mm | None (true) | None – this image acquired in the same beam scan |
| 20 kV FSE image, top diodes, camera retracted 25 mm | Affine transform (shift) | DC electromagnetic field shift from camera movement |
| 20 kV FSE image, top + bottom centre diodes, camera fully inserted | Projective transform (tilt) | Beam-scan misalignment between SEM accelerating voltages |
| 10 kV FSE image, top + bottom centre diodes, camera fully inserted | n/a (true) | n/a |

Figure 5 shows the images after alignment using TrueEBSD. Black regions bordering each image show the extent of image displacements in the alignment step. The pixel positions of the EBSD map and all images could be directly overlaid after cropping the irregular image borders.

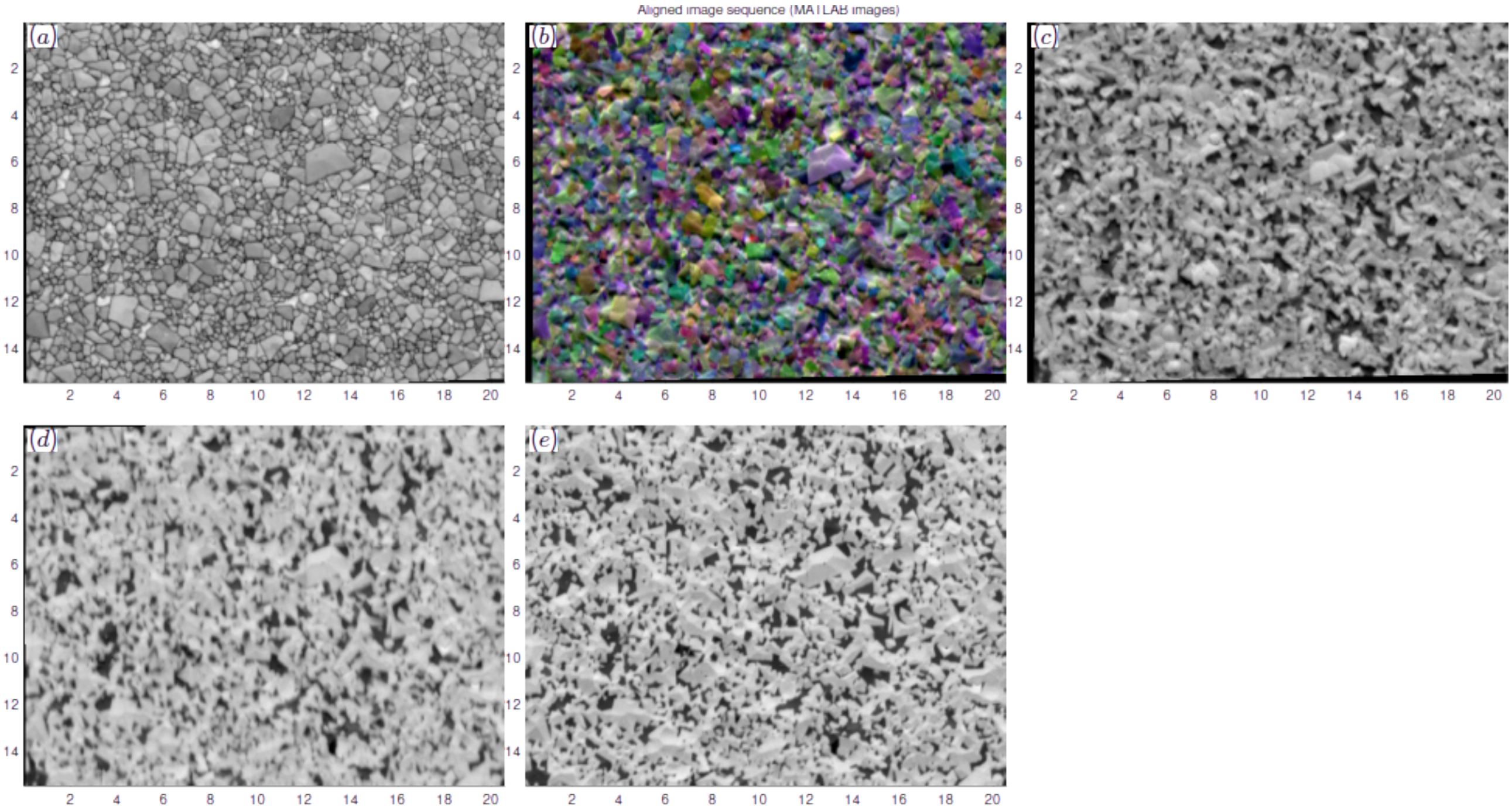

*Figure 5: Image sequence after alignment. Coordinate axes show image scales in μm.*

## 3.4 Results

EBSD maps were augmented with Co phase assignments from binary segmentation of the corresponding 10 kV FSE image after alignment using TrueEBSD. This improves the Co phase fraction estimations, but removes all Co crystallographic phase or orientation information from EBSD.

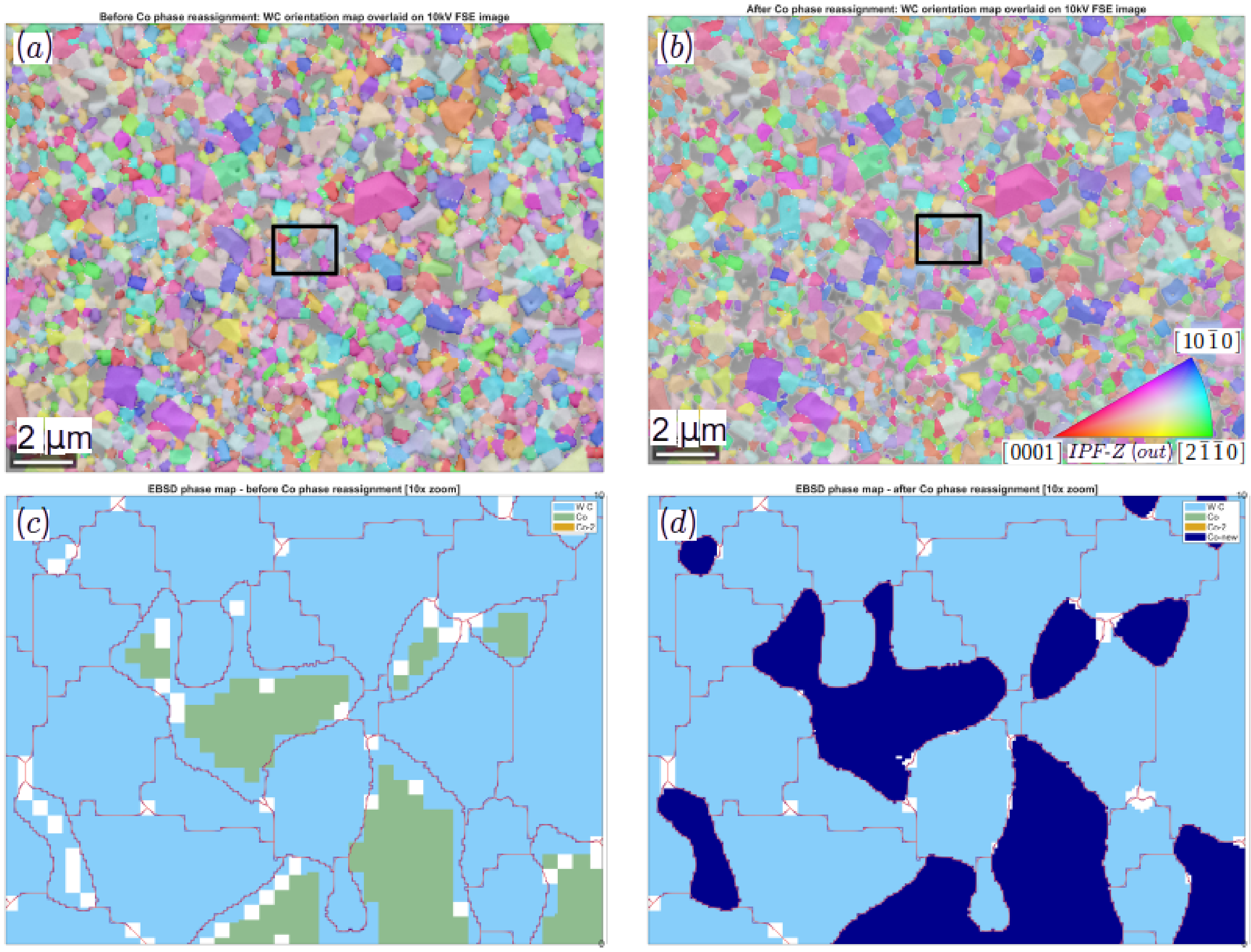

*Figure 6: (a, b) EBSD maps showing WC orientations coloured using inverse pole figure (IPF)-Z (out of screen) directions overlaid on band contrast maps. Black box outlines show zoomed-in region for (c) and (d). (a) TrueEBSD processed map; (b) after augmentation, where some WC map points have been reassigned to Co phase. (c, d) 10× zoomed-in EBSD maps, coloured by phase: WC = light blue, FCC-Co = green, Co from FSE image = dark blue. GBs from the augmented EBSD data are overlaid on both maps. (c) TrueEBSD processed map, (d) Co phase augmented EBSD map.*

WC contiguity is defined as the fraction of a WC grain's surface area that is shared by another WC grain. In EBSD, WC grain surface area is measured as WC grain surface length, and WC contiguity of each grain can be calculated directly using Equation 1.

$$C=\frac{S_{WC-WC}}{S_{WC-WC}+S_{WC-Co}} \tag{1}$$

For images, the (boundary length-weighted) average WC grain contiguity can be calculated indirectly from linear intercept lengths using Equation 2. Computing contiguity from linear intercepts drawn on BSE images is the existing state of the art method.

$$C=\frac{2N_{WC/WC}}{2N_{WC/WC}+N_{WC/Co}} \tag{2}$$

Table 3 compares average WC contiguity for the four microstructures, calculated directly from boundary lengths in the original EBSD map, the EBSD map augmented with 10 kV FSE phase assignments after TrueEBSD alignment, and from at least 200 linear intercepts drawn on BSE images of the same material.

The original EBSD map overestimated WC contiguity compared to the BSE linear intercepts method in the WC-Co grades with small features (either small WC grains, or shorter Co mean free paths due to a lower binder volume fraction), but produced a similar contiguity value in the 'High Co / Coarse WC' microstructure. This confirms that the underestimation of Co phase fraction in WC-Co composites is related to the lower and atomic-number-dependent spatial resolution of EBSD.

TrueEBSD maps augmented with phase data from FSE imaging decreased the measured contiguity for all four grades. The decrease was more pronounced in WC-Co microstructures with small features (either small WC grains, or shorter Co mean free paths due to a lower binder volume fraction).

In all four microstructures, the contiguity measured by EBSD with reassigned Co phases is smaller than measured by BSE linear intercepts. This indicates that WC grain contiguity may depend on the observation lengthscale, especially since Co binder regions are small and fill in the gaps between convex WC grains, so higher spatial resolutions enable smaller Co regions to be observed, and two WC grains sandwiching a very thin Co layer may appear to be a WC/WC boundary at coarser sampling lengthscales. Even though the spatial resolution of the BSE images and FSE-augmented EBSD maps are similar, the linear intercept method has a coarser effective spatial resolution, because it is unlikely to sample the smallest features in the BSE image, whereas EBSD grain contiguity is a full-field measurement which automatically includes all features in the EBSD map.

*Table 3: Comparison of WC contiguities calculated using different methods on the four WC-Co microstructures.*

| | **BSE linear intercepts** | **Original EBSD** | **FSE-augmented EBSD** |
|---|---|---|---|
| **High Co / Fine WC** | 0.59 | 0.69 | 0.49 |
| **Low Co / Fine WC** | 0.77 | 0.87 | 0.63 |
| **High Co / Coarse WC** | 0.47 | 0.41 | 0.33 |
| **Low Co / Coarse WC** | 0.60 | 0.68 | 0.51 |

# 4 Case study: grain boundary voids in copper

## 4.1 Context

Creep voids form at grain boundaries (GBs) in copper under certain conditions [29], [30]. TrueEBSD was used to match voids (from BSE images) to boundary types (from EBSD), and determine whether or not certain GBs or triple points (TPs) are particularly resistant to void formation.

GB voids are visible as dark pixels in the 0° tilt BSE images and can be easily distinguished by binary image segmentation, but are not distinguishable from other un-indexed regions in an EBSD map. TrueEBSD was used to align the EBSD map to a BSE image acquired using the angular backscatter detector at 0° tilt, and MTEX was used for further crystallographic analysis of preferred void nucleation sites.

Further experimental details and information about the copper polycrystal material can be found in Appendix A3 (EBSD Map ID: 29) of Reference [13], and the EBSD data presented in this work is available on Zenodo.

## 4.2 TrueEBSD workflow

Figure 7 shows the starting EBSD maps and images imported into TrueEBSD: (a) the EBSD map, (b) a false-colour FSE image (detector partially retracted, bottom three diodes), (c, d) a BSE image at 0° tilt. The EBSD map step size was 0.1 µm and the pixel size of both the FSE and BSE images was 0.05 µm. The images in (c) and (d) are the same BSE image, but pre-processed differently to highlight grain contrast and void contrast respectively; a gamma compression filter with γ=0.1 was applied to image (c) to enhance contrast in the mid-grey levels (Equation 3).

$$I_{out} = I^{\gamma} \tag{3}$$

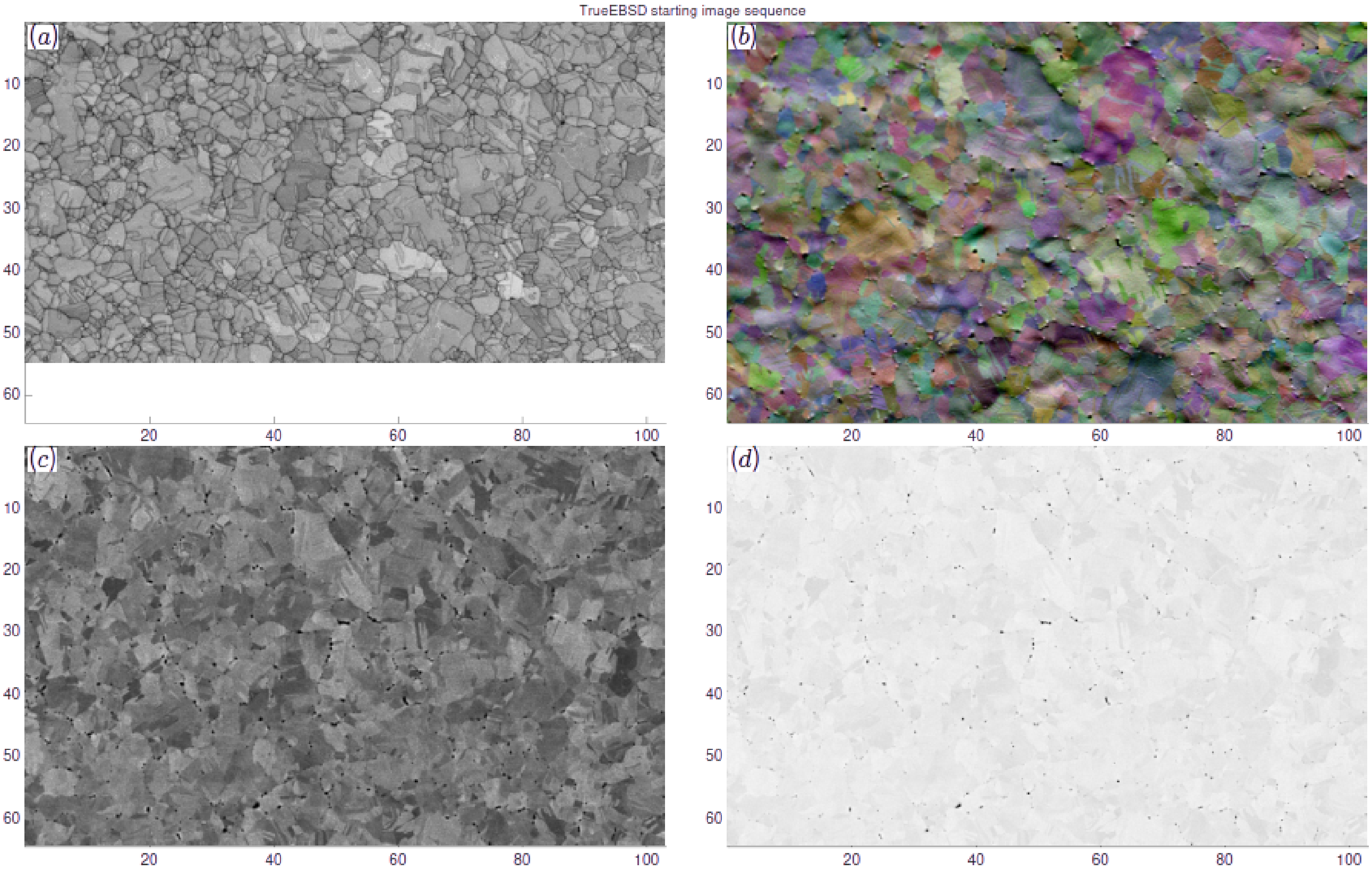

*Figure 7: Imported image sequence. (a) EBSD band contrast map, (b) false-colour FSE image, bottom three diodes, 20 kV, detector 40 mm retracted, (c) grains-contrast BSE image, 0° tilt, angular backscatter detector, γ-compression filtered (γ = 0.1), (d) topography-contrast BSE image, 0° tilt, angular backscatter detector, no γ-compression. Coordinate axes show image scales in µm.*

Figure 8 shows the same images pre-processed for image alignment. The EBSD map was aligned to the other images using the central image pixel as a reference point. The borders of the band contrast image were padded with zeros (black in Figure 8a) to match the pixel size and the EBSD map points were upsampled using nearest-neighbour interpolation to match the pixel size of the images.

Edge transforms of the EBSD band contrast map, FSE and BSE images (Figure 8 a - c) were used to generate similar features for image alignment. The EBSD band contrast image edge transform (Figure 8 a) shows mostly GB contrast, the FSE and γ-compression filtered BSE images (Figure 8 b – c) show a mixture of void and grain boundary contrast, and the BSE image with no γ-compression filter showed topography (copper / voids) contrast (Figure 8 d). No alignment was performed between the images in (c) and (d) because they were generated from the same BSE image. The distortion types between sequential image pairs are listed in Table 4.

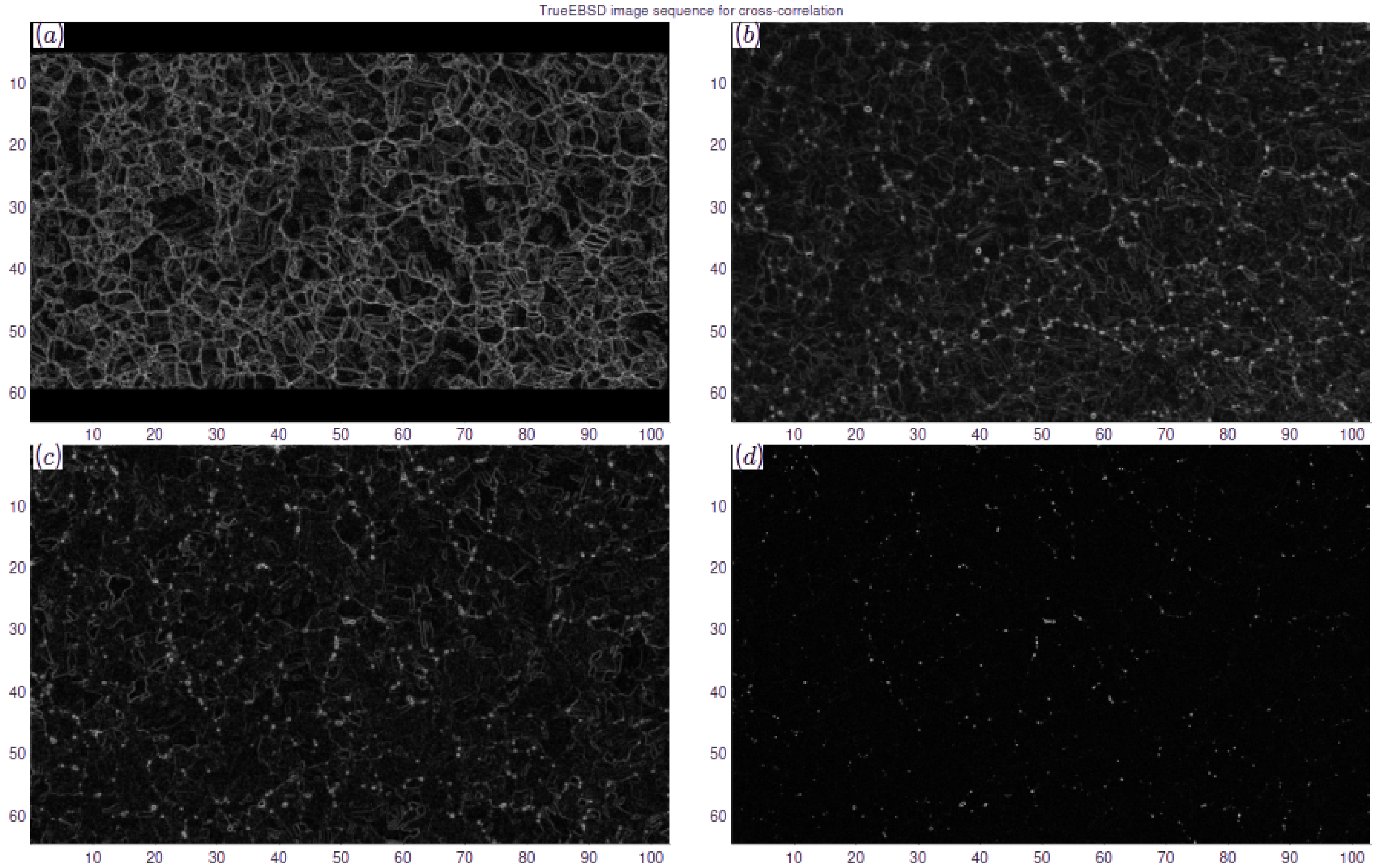

*Figure 8: Pre-processed image sequence. The edge transforms of (a) the EBSD map and (b) the FSD bottom-diodes image have shared contrast for image matching even though the images do not. Coordinate axes show image scales in $\mu$m.*

*Table 4: Physical origins of distortion types between sequential image pairs and mathematical distortion models used to fit a continuous displacement field to correct these distortions.*

| Image | Distortion model (type) | Physical origin |
|---|---|---|
| EBSD band contrast map | Affine transform + linear interpolation between rows (drift-shift) | DC electromagnetic field shift from camera movement + temporal drift |
| 20 kV FSE image, bottom diodes, camera retracted 40 mm | Projective transform (tilt) | Sample tilt |
| BSE image (γ=0.1) | None (true) | None – this image acquired in the same beam scan |
| BSE image (γ=1) | n/a (true) | n/a |

Figure 9 shows the images after alignment using TrueEBSD. Black regions bordering each image show the extent of image displacements in the alignment step. The pixel positions of the EBSD map and all images could be directly overlaid after cropping the irregular image borders.

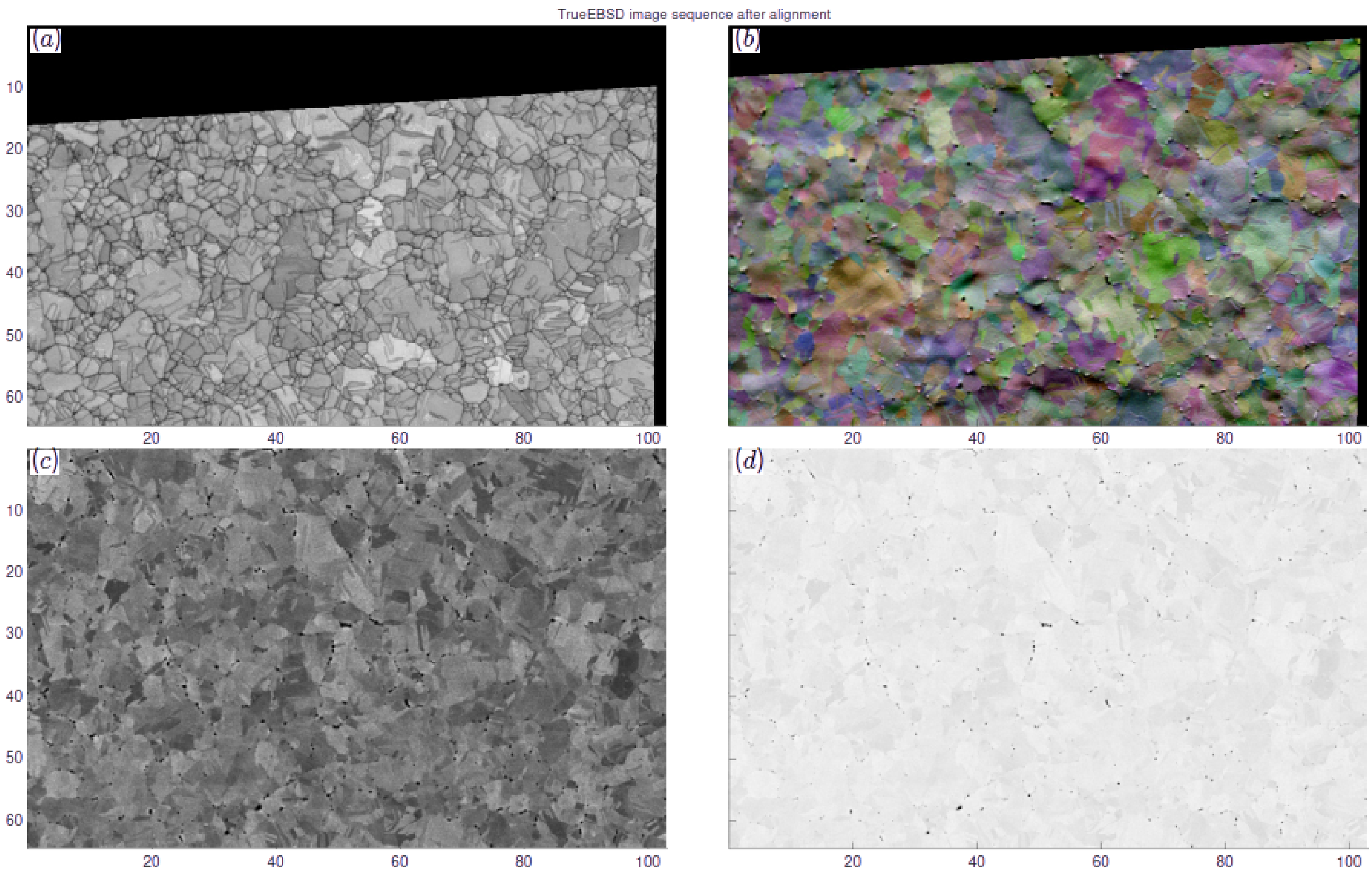

*Figure 9: Image sequence after alignment. Coordinate axes show image scales in μm.*

## 4.3 Results

Voids were identified by binary thresholding of the topography-contrast BSE image (Figure 9 d) and used to augment the aligned EBSD map (Figure 9 a) in MTEX. Points belonging to void pixels were assigned to a new phase 'voids' with arbitrary crystal symmetry and zero orientations. Figure 10 a shows the augmented EBSD map with 'Copper' and 'voids' phases. The majority of voids lie directly on a grain boundary (GB), but some voids lie a few pixels away from the boundary because of residual misalignments which were not corrected by TrueEBSD. Grains shown in Figure 10 b were reconstructed using the copper phase only, so that GBs go through voids instead of around them.

Void areas can be classified by their distance from a GB or TP. In this EBSD map, 90.2 % of void pixels are on or near a GB (including GBs attached to TPs). Many prior experimental observations on related materials [13] have shown that voids only nucleate at grain boundaries, and therefore the 10 % of void pixels which appear to lie far from a GB are not physically valid; they could be image artefacts or features that appear dark in the BSE image but that are actually not GB voids, voids that end up far from its grain boundary due to an image alignment error, or the far edges of larger voids which extend more than 5 pixels from the nearest grain boundary. 50 % of void pixels are either on or near a TP, and 50 % are far from a TP.

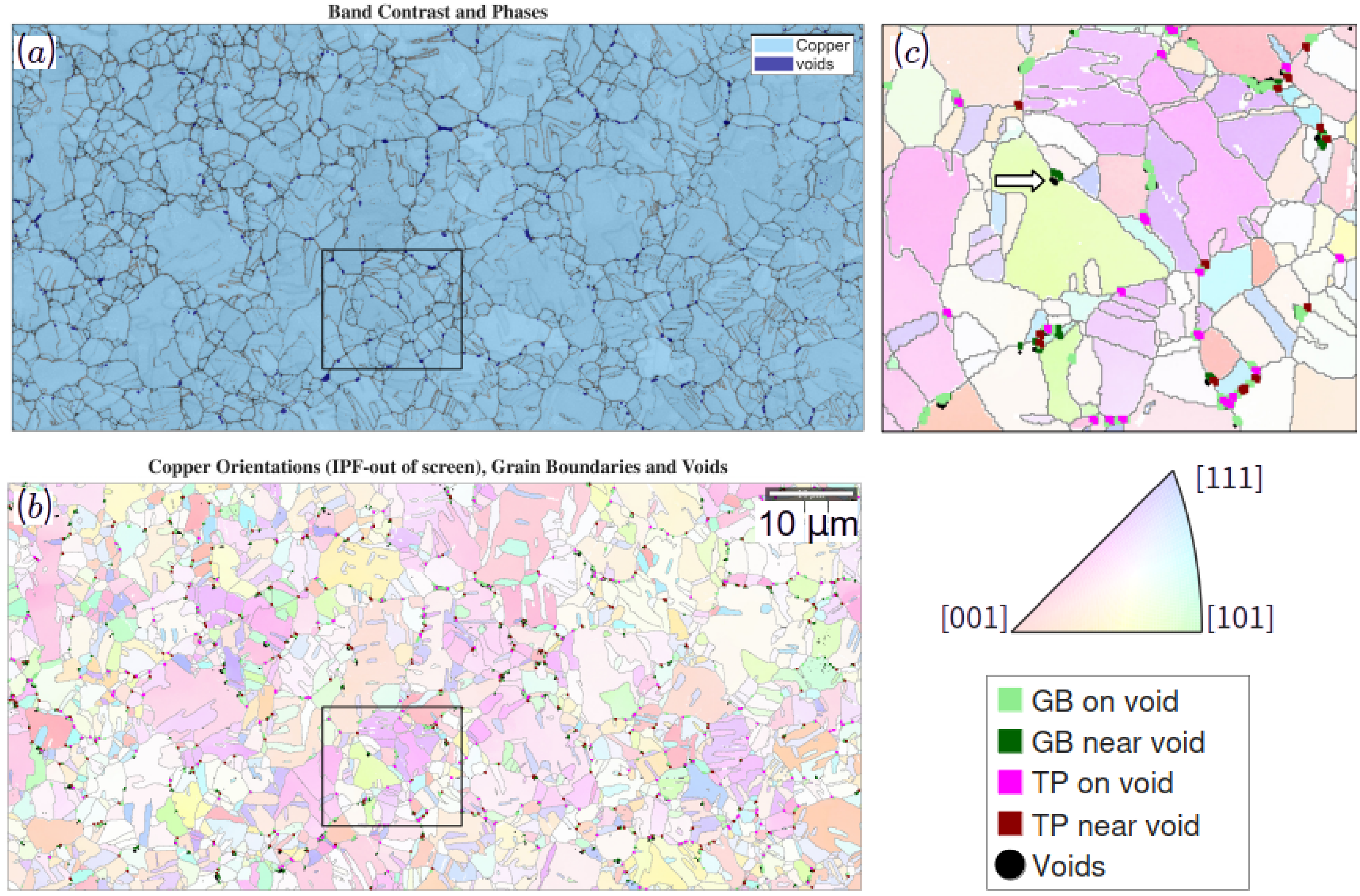

*Figure 10: GB voids analysis using augmented EBSD map after TrueEBSD alignment. (a) EBSD phase map overlaid on EBSD band contrast map (greyscale); (b) EBSD orientation map showing inverse pole figure directions pointing out of screen, copper grain boundaries in grey, and voids in black. Grain boundary segments and triple junctions near or on a void are marked in green or purple. (c) Zoomed-in copy of (b) showing region in black box.*

To accommodate residual errors after TrueEBSD alignment, GB segments and TPs within a threshold distance of a void were counted as 'near' a void. The 95$^{th}$ percentile of the TrueEBSD fit residuals summed in quadrature, which was 5 pixels for this dataset, was chosen as the threshold value. GB segments on / near a void are highlighted in Figure 10 b in light / dark green respectively, and TPs on / near a void are highlighted in light / dark purple respectively. The annotation arrow in Figure 10 c indicates a void which passes near but does not intersect a GB because of an alignment error. Even so, the GB segments near this void could be identified and are shown marked in dark green.

The crystallography of GBs and TPs near/on voids were analysed by comparing their misorientation distribution statistics. Figure 11 shows misorientation distribution axes (a – d) and angles (e) of the boundary segments which were classified as near or on a GB / TP. This microstructure contains a high frequency of [111] / 60° boundaries (Figure 11 a, e blue plot) compared to a uniform boundary misorientation distribution (Figure 11 d, e green plot). This misorientation corresponds to the Σ3 twin boundaries in FCC copper. This peak is missing from the void GBs (Figure 11 b, e orange plot), which indicates that Σ3 twin boundaries are particularly resistant to void formation. However, the misorientation distribution function of void TPs (Figure 11 c, e yellow plot) are similar to the underlying microstructure, which indicates that there are no triple junctions types that are particularly resistant to void formation in this material.

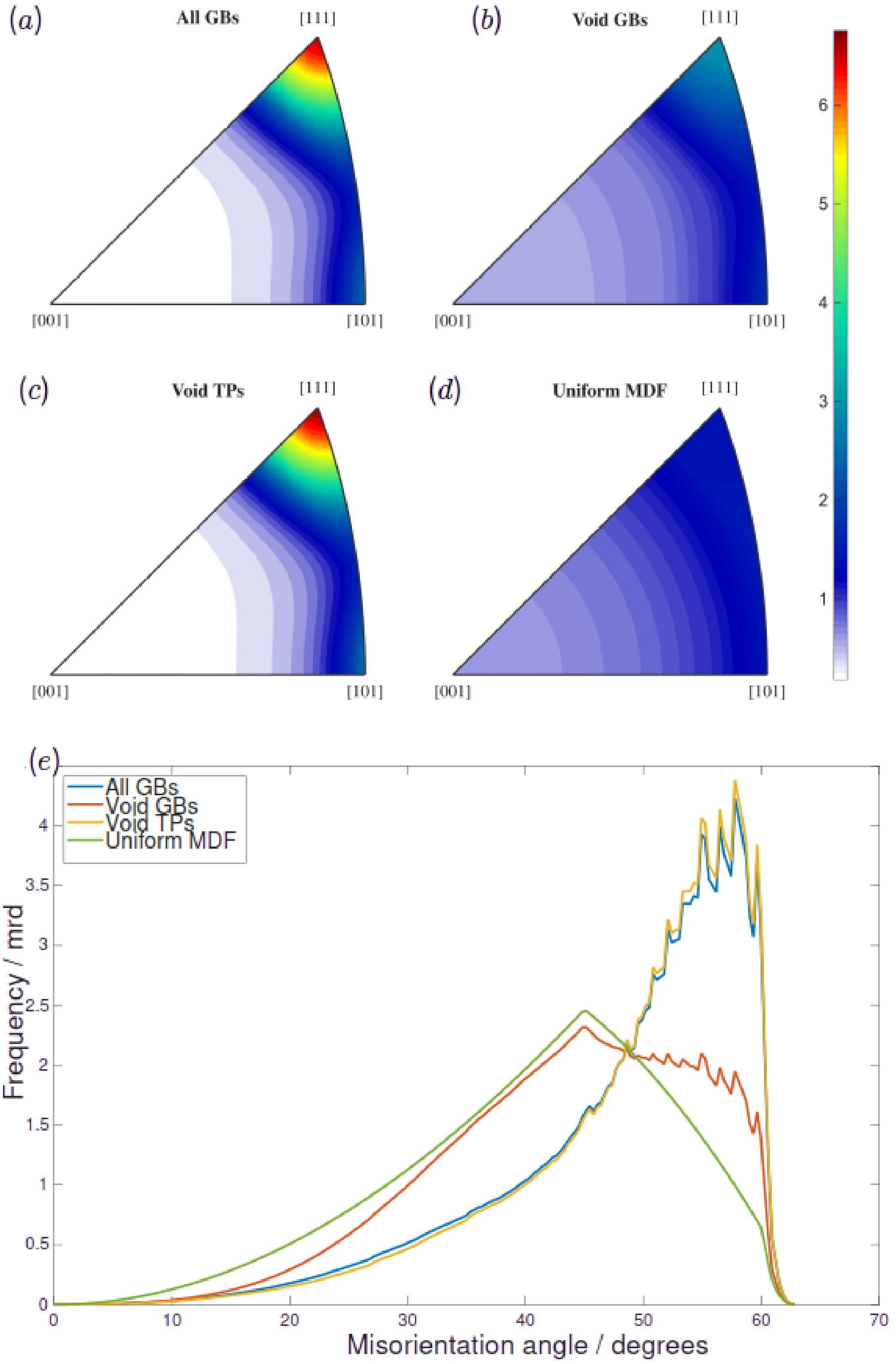

*Figure 11: Misorientation axis distributions of (a) all boundaries in the EBSD map, (b) GBs on or near a void, (c) TPs on or near a void, (d) theoretical misorientation axis distribution from a uniform orientation distribution for FCC copper, (e) corresponding misorientation angle distribution plots.*

# 5 Discussion

## 5.1 Improvements to code flexibility

The original algorithm and code prescribed the use of a specific set of images (EBSD grain boundary skeleton, near field-FSE, far field-FSE, 0° tilt BSE) which was optimised for a narrow combination of use case, available hardware, and sample types. The new algorithm is more flexible, and allows different combinations of images to be added and omitted where required.

For example, the near-field FSE image was omitted entirely from the copper EBSD map analysed in Section 4, because its contrast was dominated by short-range topography features which were visible in neither the EBSD map nor the BSE image, and most likely caused the original TrueEBSD program to fail when used in Reference [13].

Additional images can be added where required, including images with defined as having zero relative displacements (distortion type 'true'). In the copper case study (Section 4), two versions of the same copper BSE image with different image filters could be used: one with enhanced grain boundary contrast, one with enhanced voids (topography) contrast (Figure 7 c, d). In the WC-Co case study (Section 3), this links the lower-diodes false colour FSE image which shows mostly WC grain contrast, to the upper-diodes FSE image (Figure 3 b, c) which shows mostly WC/Co phase contrast, since they were acquired from the same SEM scan.

The TrueEBSD class structure, described in Section 2.2.2, allows code related to generic functions to be handled separately to a particular experimental dataset, so that TrueEBSD functions can be reused for a wide range of non-standard use cases without changing the underlying code.

For example, significant modifications to the as-supplied main script were required to run the original TrueEBSD program for aligning two Zr EBSD maps before and after SEM *in-situ* tensile deformation [1], [10], but the new TrueEBSD program can be used to run the same dataset with no code changes to the TrueEBSD functions (and if the GUI is used to import data, no coding at all is required from the user).

Small tweaks to the workflow are also possible with minimal code changes: for example, we have replaced the binary grain boundary skeleton with the band contrast map (or similar alternatives from other EBSD systems, such as pattern quality or image quality maps) as the default image for TrueEBSD cross-correlation. The band contrast map is more robust for image matching because it is a greyscale image which offers greater information depth than a binary skeleton image, does not depend on a correct EBSD indexing solution, and provides the required image contrast (different brightnesses in different grains, and grain boundaries appear dark).

TrueEBSD now uses MTEX to store EBSD data, and therefore benefits from data import and post-processing tools available in MTEX, including handling of different reference frames. For example, TrueEBSD was used to overlay multiple EBSD maps from three different SEM-EBSD systems, to compare absolute orientation measurement uncertainty in Reference [31]. This required compositing data from multiple SEM and EBSD manufacturers which are stored in different data formats and use different reference frame conventions. Users can also benefit from MTEX's range of EBSD and grain analysis tools to analyse their data after TrueEBSD processing, such as boundary length statistics shown in the WC contiguity case study (Section 3) and misorientation distribution functions in the copper voids case study (Section 4).

## 5.2 Graphical user interface

The main advantage of the GUI (described in Section 2.2.3) is that loading a dataset for the first time is simpler, especially when loading images stored within HD5F data containers which may have non-descriptive name labels (such as "Electron_Image_14", Figure 1 a). A GUI also enables TrueEBSD to be compiled and run as a standalone MATLAB application, since an app user cannot edit the TrueEBSD script.

The main disadvantage of a graphical user interface is that re-running the program requires the user to repeat all the menu selections with the same input parameters, whereas in a script interface the user would only need to run the script. All input parameters are stored in the output @trueEbsd object in both cases, but they are more easily accessible through a script.

## 5.3 Surface topography and tilt distortion

BSE images acquired from an un-tilted sample at lower electron energies (e.g. 5 kV) would be an even better reference image for WC contiguity calculation than the 10 kV FSE images used the WC-Co case study (Section 3), since SEM image tilt distortions are significant at the 70° tilt required for EBSD mapping. However, these samples have a surface topography of ~ 2 μm dimples due to argon broad-ion-beam milling. Figure 12 shows that this topography causes the local sample surface tilt to vary by a few degrees, and leads to significant and sharply varying distortions at 70° tilt for EBSD. Tilt correction using a single projective transformation assumes that the sample surface is planar.

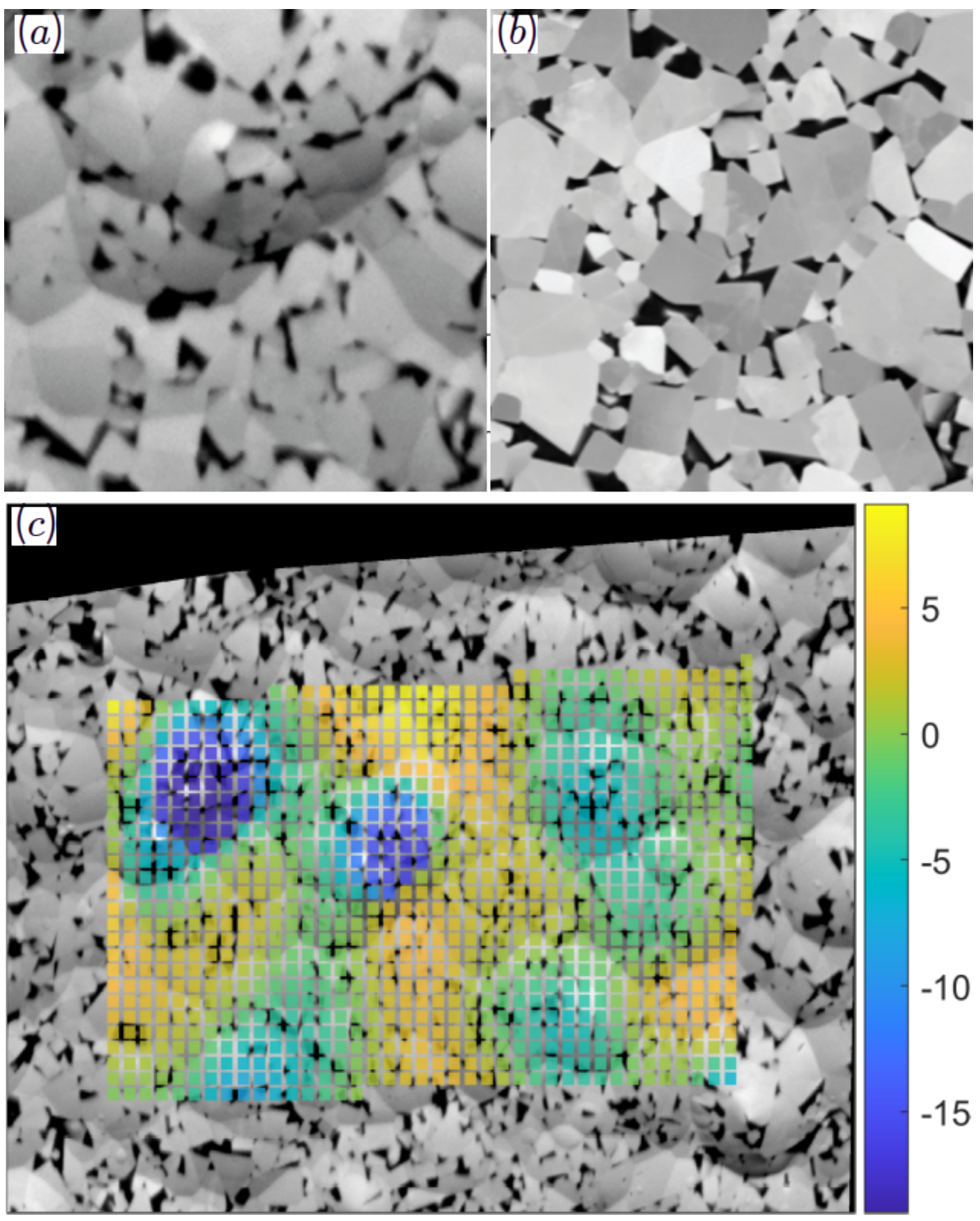

*Figure 12: Sharply varying displacement field caused by tilt distortions on an ion beam polished surface, which could not be adequately corrected using TrueEBSD. (a) FSE image (70° tilt), (b) BSE image (0° tilt), (c) Measured local displacements / pixels.*

Although Figure 12 c shows that TrueEBSD could measure these short-range distortions near the centre of the field of view, it could not adequately correct them, because they could not be extrapolated to the image edges. Cross-correlation ROI could not be placed on these locations because they would go off the edge of the image. Similar short-range spatial distortions were observed by Winiarski et al. [14] in 3D-EBSD maps serially sectioned by broad ion beam milling, and correcting these distortions required manual control point selection.

In TrueEBSD, the penalty of a fully-automated process is that the alignment near image edges is often poorer, especially if it is not appropriate to extrapolate the analytical distortion model from measured displacements near the image centre. This also applies to correcting SEM image drift, and is the main reason why automatic iterative selection of the smallest viable ROI size, described in Section 2.2.5, was important for ensuring a robust solution.

# 6 Conclusions

The TrueEBSD image alignment and distortion correction tool has been reimplemented as an MTEX add-on, using a a modular class-based structure to enable easy customisation for a wide range of applications. Other new features include a GUI data loader, computational speed-up using MEX files, and automatic optimisation of cross-correlation ROI sizes.

The copper EBSD map could be augmented with a 'voids' phase by combining the EBSD map orientations with SEM-BSE image data. Statistical analysis of grain boundary void locations in the copper polycrystal showed that Σ3 twin boundaries were particularly resistant to void formation, but no particular triple points were particularly susceptible or resistant to void formation.

The WC-Co EBSD map could be augmented with more accurate Co phase information by combining the EBSD map, which typically underestimates Co fraction, with a 10 kV top-diodes FSE image. This provided a more accurate measure of Co phase fraction, although crystallographic information related to the Co phase was lost in this process. This enabled WC grain contiguities, which is related to the length ratio of WC/Co and WC/WC boundaries, to be measured from the EBSD map.

# 7 Data and source code availability

The source code for TrueEBSD is available at https://github.com/vtvivian/mtex-trueEbsd. The latest release, used for drafting this paper, is v2.1.

Data for the copper case study are available on Zenodo https://zenodo.org/records/16902083.

Data for the WC-Co case study presented in this work are available upon request. A different example WC-Co dataset, also suitable for TrueEBSD and WC contiguity analysis, is available on Zenodo: https://zenodo.org/records/13870131.

# 8 Author contributions: CRediT

Vivian Tong: conceptualization, formal analysis, funding acquisition, investigation, software, writing – original draft, visualisation;

Stefan Olovsjö: investigation, methodology, resources;

Rachid M'Saoubi: investigation, methodology, resources;

Mathias Grabner: investigation;

Manuel Petersmann: investigation, supervision;

Liam Wright: validation, software.

# 9 Acknowledgements

This work was funded by an NPL Explorers award (2024). Alex Shard (National Phyical Laboratory UK) is thanked for helpful discussions during the NPL Explorers Award project. Ralf Hielscher (TU Bergakademie Freiberg) and Rüdiger Killian (Martin Luther Uni. Halle) are thanked for MTEX support.